\documentclass{aa}

\usepackage{graphicx}
\usepackage{txfonts}
\def\gv{2002\,GV\ensuremath{_{31}}}

\def\jjs{2007\,JJ\ensuremath{_{43}}}
\def\ortiz{2007\,OR\ensuremath{_{10}}}
\def\eg{{\it e.g.\ }}
\def\ie{{\it i.e.\ }}

\begin{document}

   \title{Uninterrupted optical light curves of main-belt asteroids from the K2 Mission}



   \author{R. Szabó
          \inst{1}
          \and
          A. Pál\inst{1,2}
          \and
          K. Sárneczky\inst{1,3}
          \and
          Gy. M. Szabó\inst{1,3,4}
          \and
          L. Molnár\inst{1}
          \and
          L. L. Kiss\inst{1,3,5}
          \and
          O. Hanyecz\inst{1,2}
          \and
          E. Plachy\inst{1}
          \and
          Cs. Kiss\inst{1}
          }

   \institute{Konkoly Observatory, Research Centre for Astronomy and Earth Sciences, Hungarian Academy of Sciences, H-1121 Budapest, Konkoly Thege Mikl\'os \'ut 15-17, Hungary\\
              \email{rszabo@konkoly.hu}
         \and
             E\"otv\"os Lor\'and Tudom\'anyegyetem, H-1117 P\'azm\'any P\'eter s\'et\'any 1/A, Budapest, Hungary
                     \and 
                     Gothard-Lend\"ulet Research Team, H-9704 Szombathely, Szent Imre herceg \'ut 112, Hungary 
          \and 
                     ELTE Gothard Astrophysical Observatory, H-9704 Szombathely, Szent Imre herceg \'ut 112, Hungary
                     \and 
                     Sydney Institute for Astronomy, School of Physics A28, University of Sydney, NSW 2006, Australia      
             }

   \date{received; accepted}

 
  \abstract
   {Due to the failure of the second reaction wheel, a new mission was conceived for the otherwise healthy \textit{Kepler} space telescope. In the course of the K2 Mission, the telescope is staring at the plane of the Ecliptic, hence thousands of Solar System bodies cross the K2 fields, usually causing extra noise in the highly accurate photometric data. }
   {In this paper we follow the {\it someone's noise is another one's signal} principle and investigate the possibility of deriving continuous asteroid light curves, that has been unprecedented to date. In general, we are interested in the  photometric precision that the K2 Mission can deliver on moving Solar System bodies. In particular, we investigate space photometric optical light curves of main-belt asteroids.}
   {We study the K2 superstamps covering the M35 and Neptune/Nereid fields observed in the long cadence (29.4-min sampling) mode. Asteroid light curves are generated by applying elongated apertures. We use the Lomb-Scargle method to find periodicities due to rotation.}
   {We derived K2 light curves of 924 main-belt asteroids in the M35 field, and 96 in the path of Neptune and Nereid. The light curves are quasi-continuous and several days long. K2 observations are sensitive to longer rotational periods than usual ground-based surveys. Rotational periods are derived for 26 main-belt asteroids for the first time. The asteroid sample is dominated by faint (>20~mag) objects. Due to the faintness of the asteroids and the high density of stars in the M35 field, only 4.0\% of the asteroids with at least 12 data points show clear periodicities or trend signalling a long rotational period, as opposed to 15.9\% in the less crowded Neptune field. We found that the duty cycle of the observations had to reach $\sim$ 60\% in order to successfully recover rotational periods.}
   {}


   \maketitle
%

\section{Introduction}

\begin{figure*}
\centering
\includegraphics[width=18.4cm]{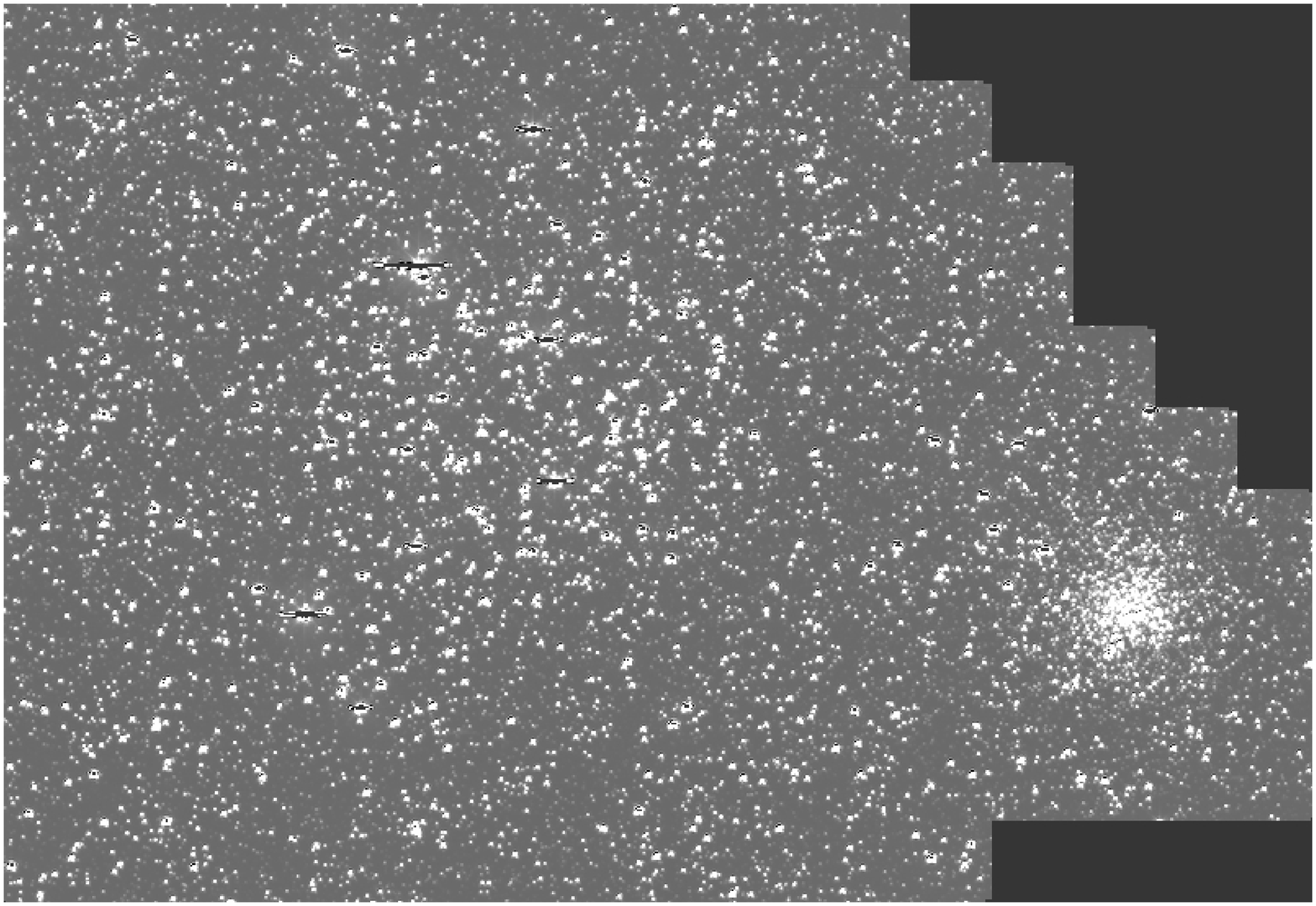}
\caption{Mosaic of the M35 open cluster (as well as NGC~2158) as seen by K2. The area is covered by 154 sub-apertures amounting to 800$\times$550 pixels ($53^\prime\times36^\prime$).}
\label{FigM35}
\end{figure*}

The \textit{Kepler} space telescope revolutionized time-domain astronomy, and its unique capabilities were demonstrated by the detection of short \citep{sanchis2014} and long period transiting exoplanets \citep{kipping2016}, the application of stellar seismology \citep{chaplin2011}, and the renewed interest in studying classical variable stars \citep{gilliland2010}. In the latter case \eg a new dynamical phenomenon was discovered in RR\,Lyrae stars \citep{szabo2010}, whose detection had been previously hampered by the diurnal variations affecting ground-based observations. 

In 1901 von Oppolzer first noticed the brightness variation of an asteroid, (433) Eros \citep{oppolzer1901},  and its first correct period was published in \citet{bailey1913} with several other minor planets.
During the more than hundred years since then the light variation of over 10000 main-belt asteroids were measured, but the length of the continuous observations has been always limited by the maximal duration of a winter night. 
The re-purposed \textit{Kepler} mission (K2) \citep{howell2014} made it possible for the first time to measure the brightness of a large number of main-belt asteroids quasi-continuously. In this paper we show the results obtained from the photometry of  close to 1000 main-belt asteroids,  most of them followed continuously for up to 3--4 days or longer, a small fraction of them up to 6 days in two large super-stamps of the K2 Mission. 

In a series of works we have been investigating the possibilities of high-precision space photometric observations of Solar System objects with the rejuvenated \textit{Kepler} space telescope \citep{howell2014}. In \citet{szabo2015} the effects of main-belt asteroid encounters on K2 photometry of stellar targets were investigated. In \cite{pal2015} we analyzed two faint Trans-Neptunian Objects, namely \gv and \jjs, measuring their rotation periods. These are among the faintest objects \textit{Kepler} has measured so far.  We also outlined the methodology to deal with these moving targets that \textit{Kepler} had not been designed for. Special masks were allocated to these targets making their continuous observations possible. By complementing recent K2 observations with archival \textit{Herschel} data, we analyzed the thermophysical parameters of \ortiz\ in \cite{pal2016}.

In this work we turn our attention to two special fields that K2 has observed, both of which have been covered by multiple sub-apertures, creating large enough fields to search for asteroids. In Campaign 0 a well-known, bright open cluster, M35 (NGC\,2168) was observed with Kepler. It was covered with a mosaic of 154 50$\times$50 pixel small stamps amounting to 800$\times$550 pixels ($53^\prime\times36^\prime$ on the sky), EPIC IDs ranging from 2000000811 to 200000964. The observed field includes the open cluster NGC~2158 as well. 
By quickly investigating the images it was immediately obvious that hundreds of asteroids crossed this field of view. We choose these observations because of the large, contiguous field and the large number of asteroids available. Campaign 0 covers the period Mar 8 to May 30, 2014, it was implemented as a full-length engineering test to prove that K2 was a viable mission. The \textit{Kepler} spacecraft was not in fine point for the first 16 days of C0, causing large photometric scatter. Eventually, the \textit{Kepler} spacecraft went into safe mode that lasted for about 24 days. After that stopping, high-quality, fine-point measurements began, which span 35 days. The data quality improved in the second half of the campaign. We used data from only this part of the campaign. Jupiter was crossing the field, but fell on the dead Module\,3, and caused only increased background flux.

\begin{figure*}
\centering
\includegraphics[width=18.4cm]{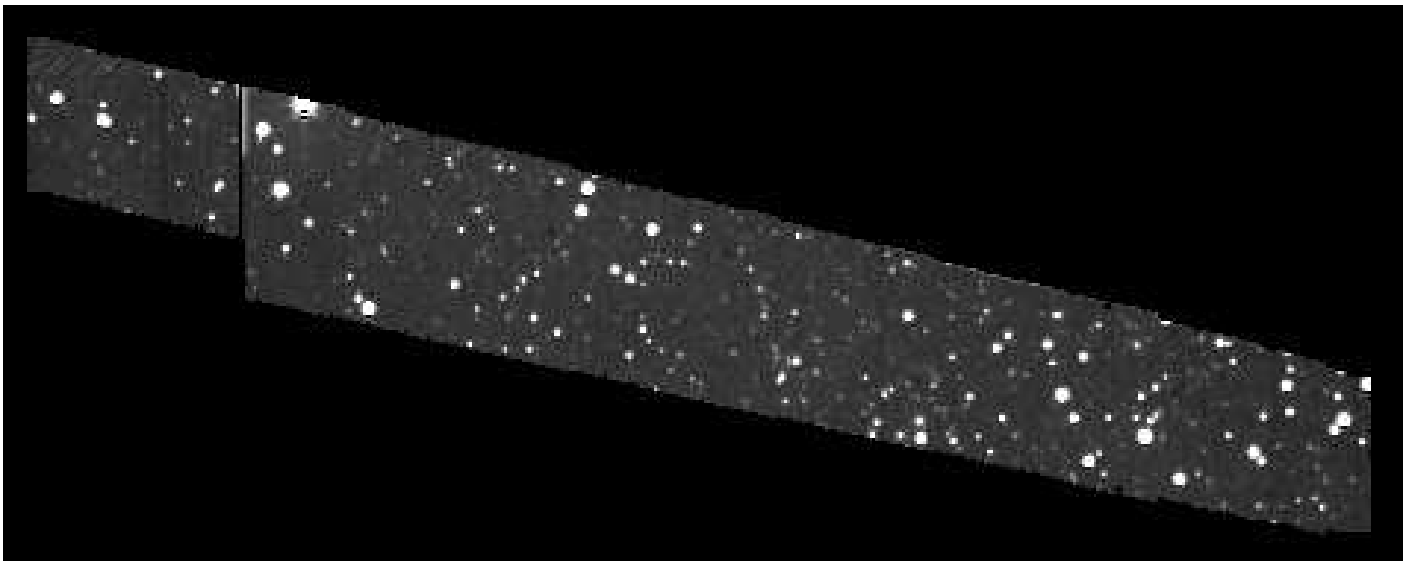}
\caption{Mosaic covering the paths of Neptune and Nereid observed by the K2 Mission. The length of the mosaic is approximately 20$^\prime$.}
\label{FigNereid}
\end{figure*}

Campaign 3 started on 15 November, 2014, and ended on 23 January, 2015. 
In this campaign Neptune and its satellite, Nereid was observed (GO IDs: 3057, 3060, 3115), their path was also tiled with 305 narrow strips of pixel masks (EPIC IDs 200004468--200004762). We refer to this field as Nereid field henceforth. While Campaign 0, and especially the vicinity of M35 contains a crowded field, this small stripe of the sky gave an opportunity to analyze light curves of asteroids usually free of too large number of stellar sources. However, due to the proximity of Neptune we had to deal with other problems (high background, saturation, etc). The change in bandwidth for pointing control (from 50 to 20 seconds) for C3 resulted in an increase in SNR for short cadence by a factor of roughly 4--9, with the larger improvement seen at the higher frequency end. Campaign 3 had a nominal duration of 80 days, but an actual duration of only 69.2 days. The campaign ended earlier than expected because the on-board storage filled up faster than anticipated due to unusually poor data compression\footnote{http://keplerscience.arc.nasa.gov/k2-data-release-notes.html}.

A similar study about Jovian Trojan asteroids observed by the K2 Mission will be published in a related paper (Szab\'o et al., 2016).


\section{Data analysis}

\begin{figure}
\centering
\includegraphics[width=1\columnwidth]{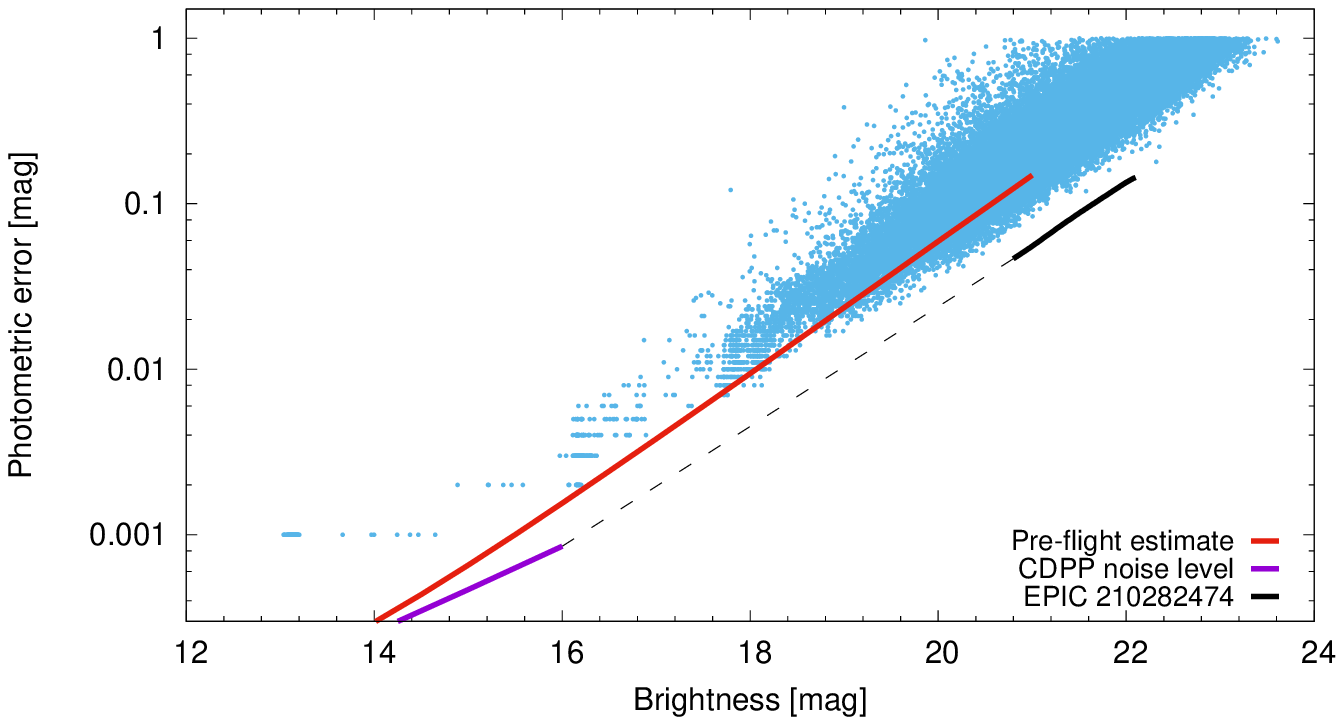}
\caption{Photometric errors of the individual K2 asteroid measurements in the M35 field, as a function of the apparent brightness of the asteroids (light blue dots). The red line is the \textit{Kepler} pre-flight estimate of the photometric precision  \citep{instrumenthandbook}\protect\footnotemark, the purple line is the scaled Combined Differential Photometric Precision data \citep{characteristics}, the black line is the photometry of a faint RR Lyrae star in Leo IV for comparison \citep{molnar2015}, the dashed line is a linear interpolation between the last two.}
\label{magerr}
\end{figure}

In Fig.~\ref{FigM35} we show the result of stitching the small sub-fields covering the open cluster M35, while Fig.~\ref{FigNereid} displays the path covered by Nereid in the vicinity of Neptune. 

\subsection{Photometry}

We used only long-cadence (29.4-min sampling) observations of both fields. 
For preparing the light curves and deriving  photometry we used the FITSH package \citep{pal2012}.
Some aspects of obtaining precise photometry of moving targets with the \textit{Kepler} space telescope during the K2 Mission have already been discussed by our group in \cite{pal2015,pal2016} and in \cite{kiss2016}. Here, due to the main belt asteroid targets we had to deal with more elongated trails during the long cadence observations as opposed to Trans-Neptunian objects in our earlier works. The method is based on a process that fits a circular aperture convolved with the apparent track of the asteroid. The track is approximated with a linear curve. The details of the methodology is given in Szabó et al. (in prep). In Fig.~\ref{magerr} we plot the photometric error of each individual data point for every asteroids in the M35 field as a function of brightness.
The error contains the noise from the background and the shot noise. The background noise is determined from a  
circular or elongated ring around the target, while the shot noise is computed from the known electron/ADU conversion rates. For the brightest targets we added a conservative upper error limit (0.001 mag for the brightest objects, 
0.002 mag for slightly fainter observations, etc.), since the error computation gives an unrealistically low error limit due 
to neglecting systematic errors (e.g. errors caused by passing through bright stellar objects). Our approach is an attempt to compensate 
for this underestimation. We believe that this choice does not affect the information content of the figure and it affects only a negligible 
portion of the targets. The precision reaches a few mmag for the brightest objects, 0\fm01 for a ${\rm 18^{th}}$ magnitude object and 0\fm1 for an asteroid of the ${\rm 20^{th}}$ magnitude. We note that the error values do not contain systematic errors. The precision we achieved is better than the conservative pre-flight estimate of \textit{Kepler}, but slightly worse than the precision derived from actual photometry of stellar targets. For the latter comparison we scaled the Combined
Differential Photometric Precision (CDPP, \citealt{characteristics}) values of the original \textit{Kepler} mission to a single long cadence exposure. However, these data only extended to 16 mag, so we also included the measurements of the extragalactic RR Lyrae EPIC 210282474, the only variable star in Leo IV that was not affected by blending \citep{molnar2015}. Finally, we added a linear interpolation between the two data sets.

 \footnotetext{Tabulated values: http://keplergo.arc.nasa.gov/CalibrationSN.shtml}
 
\begin{figure}
\centering
\includegraphics[width=9.0cm]{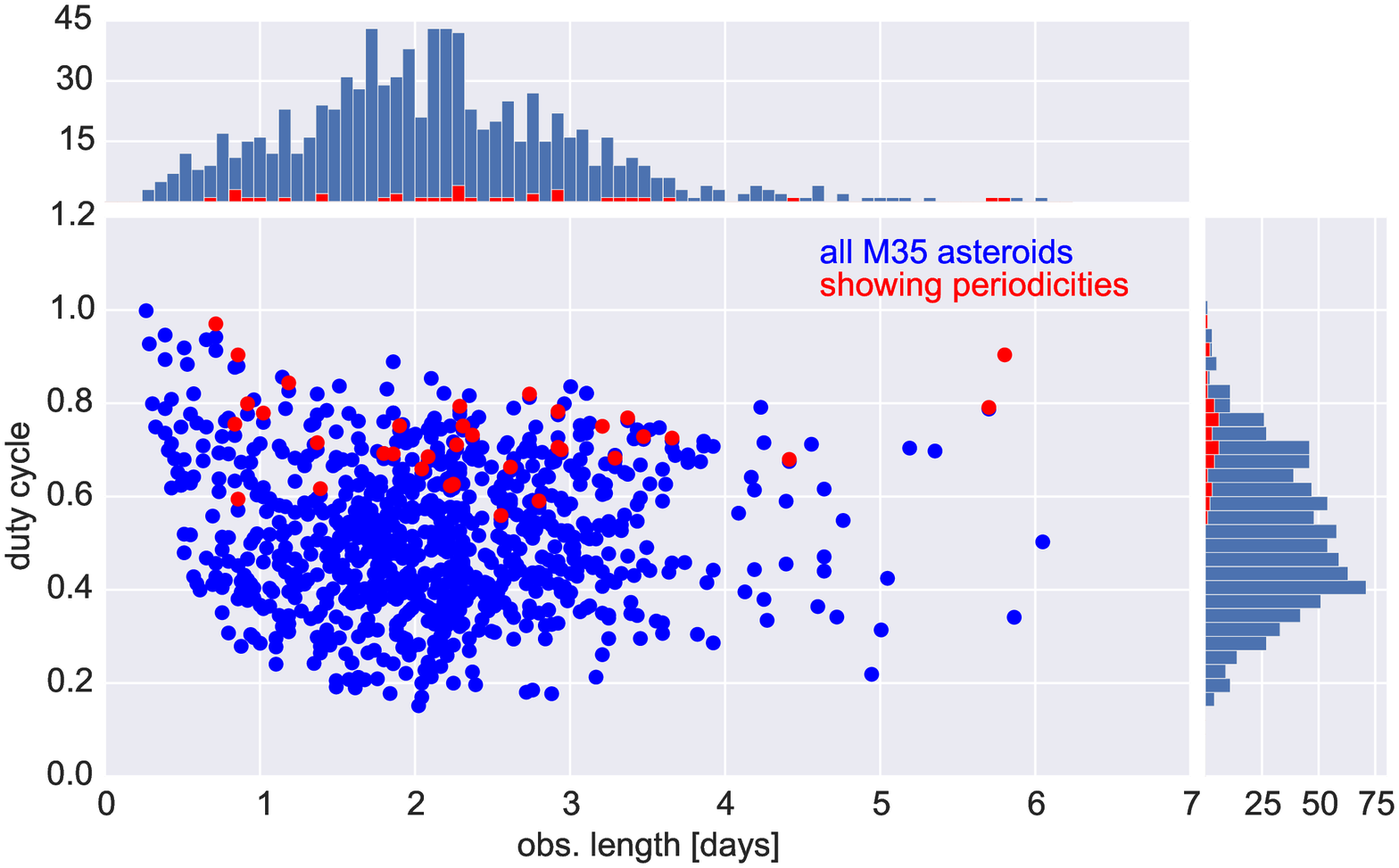}
\caption{Duty cycle vs. length of observations for asteroids in the M35 K2 field. Blue dots are all the 924 asteroids crossing the superstamp, red ones are those that exhibit  significant periodicity or trend. The histograms show the distribution of the observing length and the duty cycle.}
\label{duty}
\end{figure}

\begin{table*}
\centering
\caption{Sample table of the observed asteroids in the K2 M35 superstamp. The whole table containing 867 objects with at least 12 photometric points is available electronically.}
\begin{tabular}{ccccccccc}
\hline\hline
ID & start & end & length & \# of obs. & duty cycle & USNO R & per. & ampl. \\
& BJD - 2,456,700 & BJD - 2,456,700 & [d] &  & [0..1] & [mag] & [h] & [mag] \\
\hline
111 & 101.2871 & 103.4326  & 2.1663  & 72 & 0.679 & 13.1 & - & - \\       
228 & 104.2704 & 105.1286  & 0.8790  & 25 & 0.581 & 17.5 & 6.437  & 0.158 \\
767 & 86.3093  & 89.2313   & 2.9428  & 112 & 0.778 & 16.2 & >60. &  >0.1 \\
2462 & 78.8102 & 80.7105   & 1.9211  &  60 & 0.638 & 18.1 &  - & - \\
... & ... & ... & ... & ... & ... & ... & ... & ... \\
\hline
\end{tabular}
\label{tab1}
\end{table*}

\begin{table*}
\centering
\caption{Same as Table~\ref{tab1}, but for the asteroids in the Nereid field observed during  K2 Campaign~3. The full table containing 88 asteroids with at least 12 data points is available electronically.}
\begin{tabular}{ccccccccc}
\hline\hline
ID & start & end & length & \# of obs. & duty cycle & USNO R & per & ampl \\
& BJD - 2,456,900 & BJD - 2,456,900 & [d] &  & [0..1] & [mag] & [h] & [mag] \\
\hline      
2954 & 102.3946 & 106.1748 & 3.8010 & 131 & 0.704  & 18.0  & 4.691 & 0.174 \\
3785 & 120.2127 & 122.8078 & 2.6159 & 116 & 0.906 & 17.3 & 3.782 & 0.256 \\
9105 & 133.4332 & 134.9453 & 1.5329 &  61 & 0.813 & 19.0  & 4.769 & 0.770 \\
17747 &135.5379 & 137.3565 & 1.8394 &  76 & 0.844 & 20.3  & - & - \\
... & ... & ... & ... & ... & ... & ... & ... & ... \\
\hline
\end{tabular}
\label{tab2}
\end{table*}

\begin{figure*}
\centering
\includegraphics[width=9.2cm]{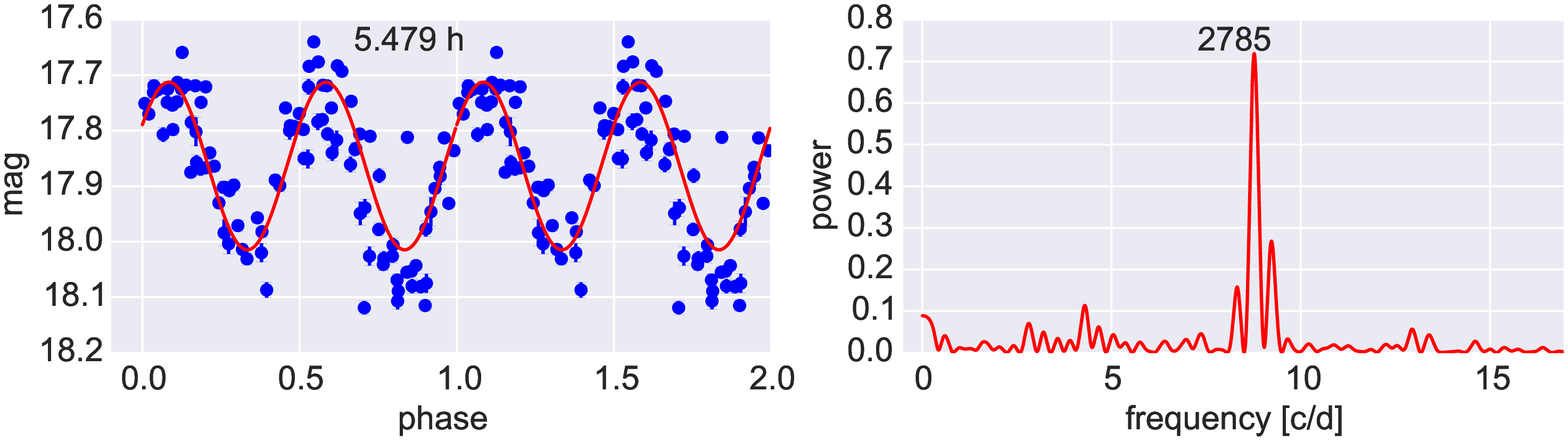}\includegraphics[width=9.2cm]{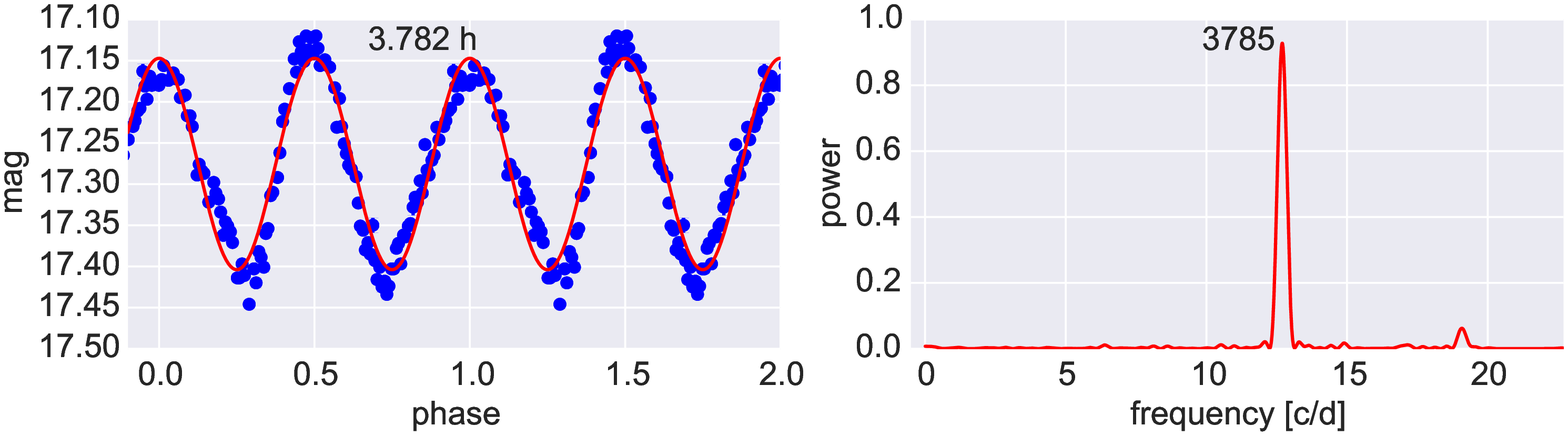}\\
\includegraphics[width=9.2cm]{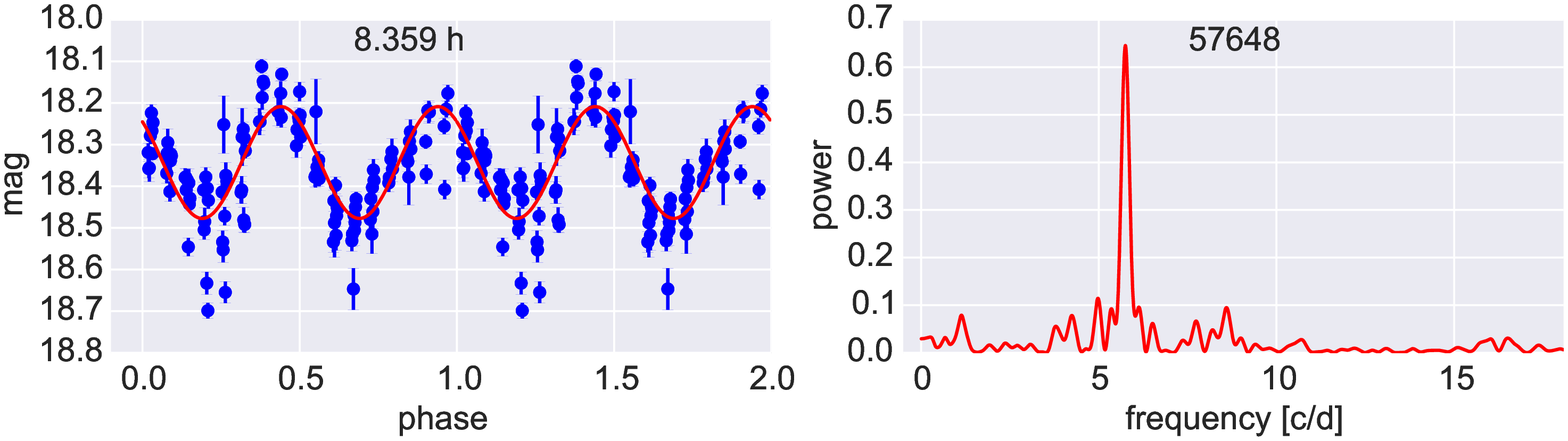}\includegraphics[width=9.2cm]{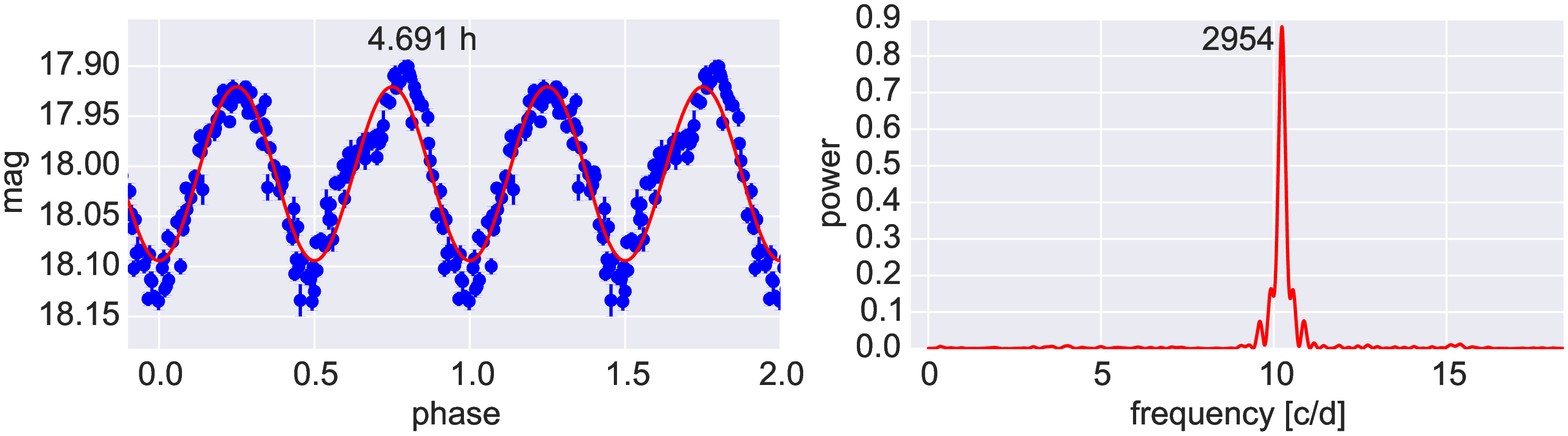}\\
\includegraphics[width=9.2cm]{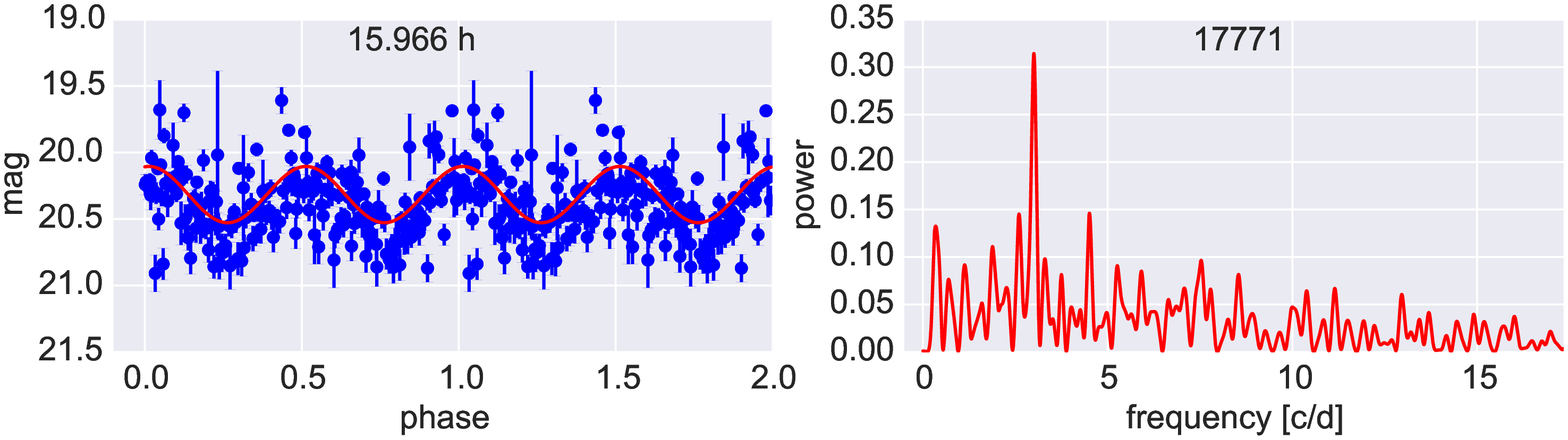}\includegraphics[width=9.2cm]{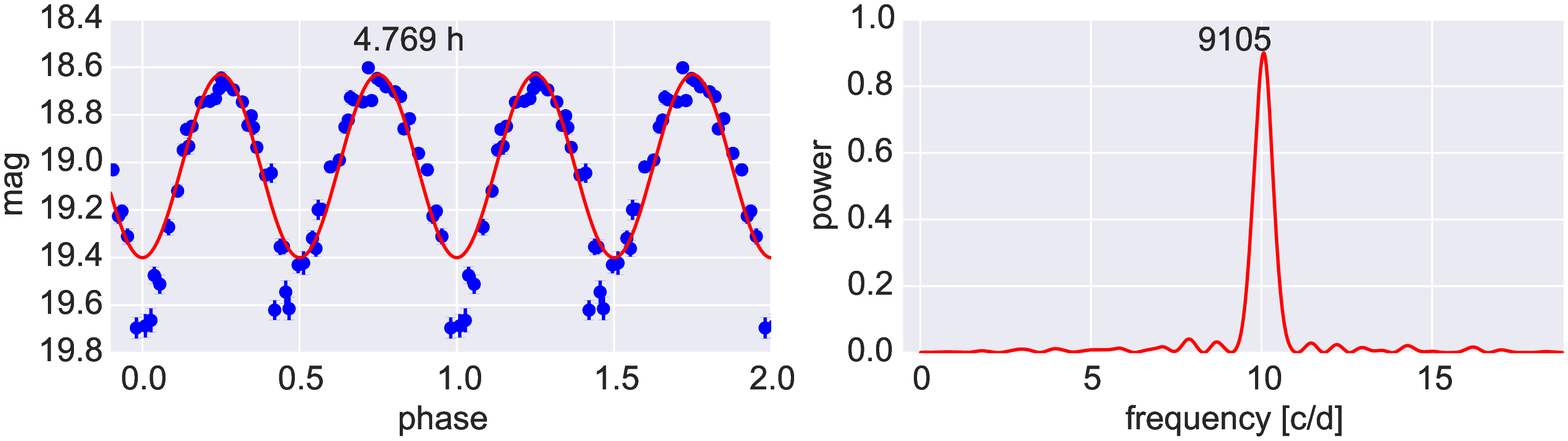}\\
\caption{Examples of periodic long cadence (29.4-min sampling) asteroid light curves in decreasing brightness order. Left panels: asteroids crossing M35 field, right panel: asteroids observed in the Neptune-Nereid path. In each case the left panel shows the K2 light curve with error bars folded by the adopted (double) period, the right panel is the Lomb-Scargle plot. The error bars for bright asteroids are smaller than the size of the symbols. The sine plotted in red is fitted to the data to guide the eye. The adopted period and the ID of the asteroid is shown in the middle of the left and right panels, respectively.}
\label{FigLC1}
\end{figure*}


\subsection{Period search}

We searched for significant periodicities using the Lomb-Scargle periodogram functions of the {\tt gatspy} Python package\footnote{https://github.com/astroML/gatspy/}. Although Fourier-based methods were considered as well, as implemented in the {\tt Period04} program package \citep{lenz2005}, we got very similar results in several test cases, therefore we decided to stick to the Lomb-Scargle method. We note that the errors of the individual photometric points are taken into account by the used implementation of the Lomb-Scargle algorithm. Only those signals were considered that were significant on the 3$\sigma$-level {compared to the background local noise in the Lomb-Scargle periodogram.} We phase-folded the light curves with the best period and its double value, then decided which gives a better fit based on a visual inspection. 
In many cases there were significant deviations in odd and even cycles. In these cases the two-period folding was chosen. If the inspection was inconclusive, or resulted in a dynamically untenable  short period (\ie less than 2 hours) then we chose the double-period solution. Following this method we did not retain any single-periodic solutions in accordance with other main-belt asteroid works.
We note that the chosen 2-hour limit is plausible, since K2 did not observe smaller asteroids in these fields. In order to demonstrate 
this let's consider a close-by asteroid of the Hungaria family in the classical Main Belt. If its brightness in opposition is 21 magnitude 
(it's fainter than 22 magnitude in the K2 field), this corresponds to H=19.0 absolute 
magnitude, which translates into a 300-m diameter assuming an albedo of 0.5. We emphasize that this is a conservative estimate.

{\bf M35 field}: 
Out of 924 asteroids crossing the M35 field 867 had more than 12 data points. We retained only these asteroids to have a set of reasonably covered light curves. Although this number might vary from light
curve to light curve depending on the distribution of data points, we found that this number gives a good coverage for the short-period asteroids 
and still gives some information for longer rotational periods.
Among the 867 light curves 23 showed clear periodicities, \ie{} above the 3$\sigma$ limit. The periods ranged from 3.89\,h to 88.41\,h, with a median of 9.83\,h. 
The length of the covered paths ranged from 0.26 days to 6.05 days with a median of 2.06 days. The median number of observations per asteroid was 48. 

{\bf Nereid field:} 
Out of 96 light curves 88 had more than 12 points in their K2 Campaign 3 light curves, of which 14 showed periodicities above the 3$\sigma$ limit.
The shortest detected period is 3.71 hours, while the longest is 14.63 hours, the median being 4.96\,h. The length of observation varied from a few data points ($\sim$ 1 hour) to 14.45 days, the latter corresponds to an asteroid that moved parallel to the apparent path of Nereid. The median length was 1.25 days. Compared to the M35 field, here we found less asteroids due to the smaller field, but the light curves are more complete, \ie less points had to be  dropped. To put it quantitatively, the median number of retained observations per asteroid was 57. This is due to the less crowded nature of the field. 

\begin{figure}
\centering
\includegraphics[width=9.0cm]{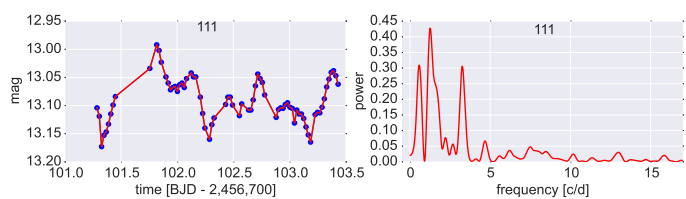}
\caption{ Light curve (left panel) and Lomb-Scargle periodogram (right panel) of (111) Ate, the brightest member of our sample. The periodogram shows three peaks, the largest one corresponds to a period of 20.63 $\pm$ 0.60\,h.}
\label{FigLC2}
\end{figure}

To end this section we show the case of (111) Ate, an example where our period search did not result in unambiguous period value. In the left panel of Fig.~\ref{FigLC2} we show the light curve. The three minima clearly define two periods in our light curve.
The periodogram (right panel of the same figure) shows three significant peaks, but we could not decide on the value based on this accurate, but short light curve. The largest peak corresponds to 20.63 $\pm$ 0.60\,h (significant at the 5$\sigma$ level), which is clearly off from the 22.072 $\pm$ 0.001\,h value found in the literature \citep{warner2009}. None of the peaks in our periodogram corresponds to the literature value. This is the brightest asteroid in our sample, 
but most probably our data set is too short to give a definite rotational period. Definitely more observations would be needed to settle the question.

\section{Results}

\subsection{Unbiased K2 main-belt asteroid sample}
In Table~\ref{tab1} we present the asteroids  identified in the M35 superstamp during the second half of Campaign 0. Only those objects are listed that showed at least 12 useful photometric data points. 
The table gives the identification, the start and end date of the time interval during which the minor planet was passing through the superstamp. The number of useful long cadence observations and the calculated duty cycle is also given. The duty cycle is 1.0 if all the photometric points were retained, and 0 if all had to be discarded. We also provide the median brightness transformed to the USNO \textit{R} band \citep{pal2015}. If reliable period and amplitude were found, these values were added to the table(s). 
In Table~\ref{tab2}. the same parameters are given for the main-belt asteroids crossing the Nereid superstamp.

As we discussed in detail in \citet{szabo2015} no (or very few) new main-belt asteroid discoveries are expected in the K2 campaign fields due to the available limiting magnitude. Indeed, among the 1020 identified asteroids in the two fields, we have not seen unknown objects. The procedure of prediction and identification of the main-belt asteroids as seen from \textit{Kepler} was described in detail in \citet{szabo2015} and was followed here, as well.

 Fig.~\ref{duty} shows the duty cycle (percentage of the number of observed 29,4-min cadences with respect to the maximally possible during the length of observation), \ie how many cadences have been lost due to technical problems originating mainly from the photometric pipeline (outliers, encountering too bright stars, stellar residuals, etc.), versus the length of observations in the case of the M35 asteroids. Most of our objects were observed for 1--4 days, and we were able to follow some of them for 5--6 days. The vast majority of our targets were observed with a duty cycle between 20 and 80\%, a handful of them above 80\%. Only one asteroid was followed with 100\% duty cycle, \ie in this case no long cadence observations had to be discarded. 
 
 Objects that exhibited significant periodicities are shown with red dots in Fig.~\ref{duty}. It is evident that the majority of the minor planets showing periodicities were found among the ones observed with high duty cycle. Namely, we found all the periodicities or long periods (seen as trends)
only where the duty cycle was close to or above 60\%. Similarly, we found that the more continuous the light curve in the Nereid field, the more the chance to detect variability at a statistically significant level.

Twelve asteroids in the M35 field exhibited long-term trends or incompletely covered half rotation periods. We identified these with longer-period asteroids. One example was found with literature data, namely asteroid (3345), for which we found a rotational period that is longer than 34 hours, and \citet{benishek2014} gives 187 hours.

In Tables~\ref{tab3} and \ref{tab4} we give the format and the content of the files that contain the K2 photometry for all the targets 
in the M35 and Nereid fields, respectively. The full tables are available online. 

\begin{table}
\centering
\caption{Photometry of main-belt asteroids observed in the M35 superstamp by K2 during Campaign~0. The full table is available electronically.}
\begin{tabular}{ccccccccc}
\hline\hline
ID & time & USNO R & R err  \\
& BJD - 2,456,700 & [mag] & [mag] \\
\hline      
111  &   101.2871  &    13.130  &    0.001 \\
111  &   101.3076  &    13.127  &    0.001 \\
111  &   101.3280  &    13.166  &    0.001 \\
111  &   101.3484  &    13.202  &    0.001 \\
111  &   101.3689  &    13.162  &    0.001 \\
...  & ...      & ...           &    ...      \\
\hline
\end{tabular}
\label{tab3}
\end{table}

\subsection{Selected light curves}

In the left panels of Fig.~\ref{FigLC1} we show a few selected main-belt asteroid light curves in the M35 field along with their Lomb-Scargle diagrams, demonstrating the high quality of K2 observations. The right panels of the same figure
 show asteroids crossing our Nereid field. Both the light curves and the periodograms are of better quality in the latter field, which underpins our suspicions that more crowding precludes obtaining high-quality light curves for main-belt asteroids with Kepler. The fact that we recovered light curves with clear periodicities, incomplete cycles or long-term trends implying long rotational period for 35 out of 867 asteroids (4.0\%) in the M35 superstamp, and 14 out of 88 objects (15.9\%) in the Nereid field also underlines this finding.

Tables~\ref{tab5}-\ref{tab6} show rotational periods and amplitudes 
along with literature data where available. In the cases where only trends or partially covered cycles were found we could establish only lower limits for the rotational period. The agreement between our rotational periods and literature values is excellent in each case despite the 
differences in observational strategies 
(multiple nights with ground-based telescopes vs.\ quasi-continuous for a few nights in the K2 mission). One exception is (111) Ate, which we discussed earlier. For more than half of our objects in Tables~\ref{tab5}-\ref{tab6} (26/47 combined) we publish unambiguous rotational parameters for the first time thanks to the continuous coverage of Kepler. The observed amplitudes are also in satisfactory agreement with available literature values. The uncertainty of the rotational period (also given in Tables~\ref{tab5}-\ref{tab6}) depends on a number of factors, such as the brightness of the object, the length of observations, and the number of observational points in the light curve.

\begin{table}
\centering
\caption{The same as Table~\ref{tab3}, but for the main-belt asteroids in the Nereid field in Campaign~3. The full table is made available electronically.}
\begin{tabular}{ccccccccc}
\hline\hline
ID & time & USNO R & R err  \\
& BJD - 2,456,900 & [mag] & [mag] \\
\hline  
2001QZ114   &    111.9575  &  19.935 & 0.069 \\
2001QZ114   &    111.9780  &  19.898 & 0.051 \\
2001QZ114   &    111.9984  &  19.983 & 0.071 \\
2001QZ114   &    112.0188  &  20.171 & 0.085 \\
2001QZ114   &    112.0393  &  20.189 & 0.079 \\
...  & ...           & ...            &    ...      \\
\hline
\end{tabular}
\label{tab4}
\end{table}

\subsection{Interesting light curves with three maxima}

\begin{figure}
\centering
\includegraphics[width=1.0\columnwidth]{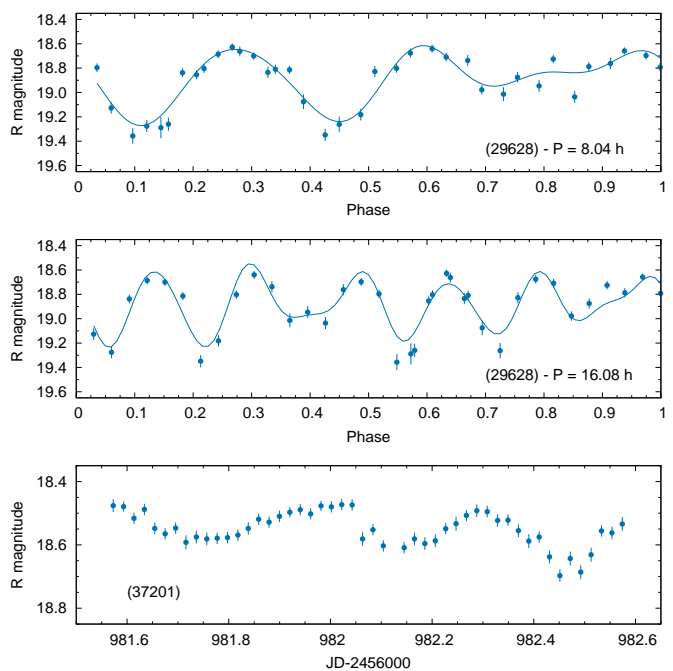}
\caption{The light curve of asteroid (29628) with three different peaks observed in the Nereid field. Upper panel: light curve folded with triple periods, lower panel: the same data folded by sixtuple period. The curve line is a sixth-order Fourier-fit. Bottom panel: the light curve of asteroid (37201).}
\label{3peaks}
\end{figure}

Trimodal (\ie three maxima during one rotation) and complex light curves are still considered as peculiar asteroid light curve shapes, although they were investigated as early as in the 1950s \citep{gehrels1962}. The first known examples of complex light curve were detected among the brightest asteroids \eg (16)~Psyche, (21)~Lutetia, (37)~Fides, (39)~Laetitia, (43)~Ariadne, (52)~Europa, (532)~Herculina  \citep{zappala1983,cellino1989}. It has also been recognized that the same asteroids exhibit bimodal or even unimodal light curves at varying phase and aspect angles, therefore, shading effects, phase effects and/or unusual rotational geometry were usually invoked to explain the complex light variations (\eg  \citet{zappala1983,michalowski1996}). More recently, \citet{harris2014} derived the higher-order harmonics due to polygonal shapes, and concluded that rotating triangles enhance the sixth harmonics, therefore leading to light curves with six maxima and not exceeding 0.156 magnitude full amplitude of the sixth harmonics. They also presented two examples: (5404)~Uemura and 2010 RC130, where the composite light curve of six maxima was far more convincing than with three maxima. Moreover, they remarked that the solution with three maxima led to short rotation periods, near the break-up barrier, which is a further argument for the longer periods and a light curve with six maxima.

Two light curves in our data were found to show this kind of complexity. Asteroid (29628) 1998~TX30 was observed in the Nereid field, during 16.7 hours. The peak-to-peak amplitude is 0\fm72, and the light curve suggests a triple or sextuple symmetry (Fig.~\ref{3peaks}). Assuming three humps, the rotational period is 8.04 hours. Another solution with 16.08 h period is also possible. The overlapping region in the latter case (P=16.08 h) extends only to five points, and despite their perfect fit to each other, drawing firm conclusions is not possible. However, the overall shape with six or eight humps is quite resembling to the examples for complex light curves in disfavour of trimodal solutions, by \citet{harris2014}. 
We note that other scenarios such as tumbling, or albedo variegation may also play a role in such light variation, but the proper investigation would require a longer data set.

We retained 50 photometric data points of asteroid (37201) 2000~WS94, covering 25 hours. The asteroid was also in the Nereid field. The light variation shows three humps with definitely incompatible shape to each other, suggesting that this asteroid also has a complex light curve (Fig.~\ref{3peaks}, bottom panel). Because of the relatively short coverage, concluding to the period of rotation is not possible. These two asteroids are therefore not included in Table~\ref{tab6}.

\begin{figure}
\centering
\includegraphics[width=8.1cm,angle=270]{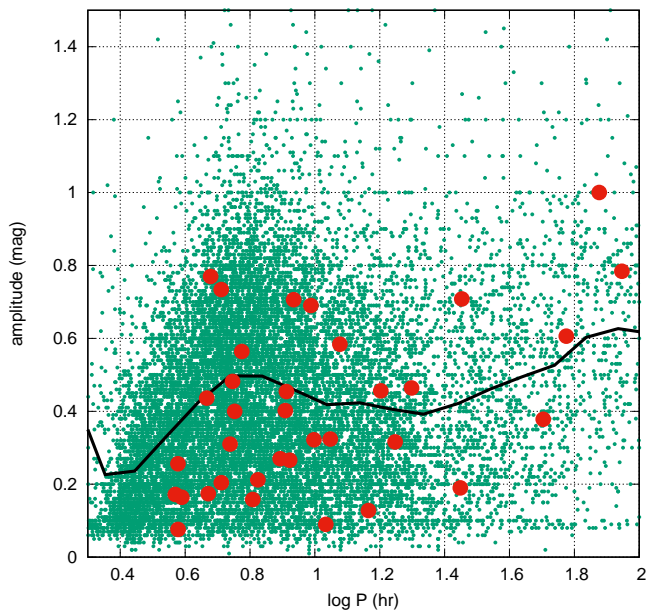}
\caption{Period-amplitude diagram of an unbiased sample of 35 asteroids from the M35 field (red points). The small green dots represent 16000 asteroids from the Asteroid Lightcurve Database \citep{warner2009}, while the thick black line shows the binned average of that large sample.}
\label{perampl}
\end{figure}

\subsection{Rotation statistics}

Fig.~\ref{perampl} shows the period-amplitude diagram of our sample of 35 asteroids with uninterrupted light curves from the M35 field. For a comparison we plotted the parameters of over 16000 asteroids taken from the Asteroid Lightcurve Database \citep{warner2009} reflecting the status of the database as of 20 February, 2016.  We plot only the rotation frequencies above the Nyquist-frequency of K2 data, where the results are comparable. The thick black line shows the binned average of the large sample. The distribution of our limited sample follows nicely that of the underlying bulk sample. 
The median rotational period in the M35 field (9.83\,h) clearly shows that we are sensitive to asteroids with relatively large periods compared to the bulk sample shown in Fig.~\ref{perampl}, whose median rotational period is 7.00\,h. This hints at the possibility that ground-based observations are usually biased toward shorter rotational periods that are easier to detect from the ground. The average period depends on the size of the asteroids (see e.g. \citet{pravec2000})  Our sample is composed by some of the smallest known asteroids, where the average rotation period is predicted to be shorter than for the largest bodies. Since we observe the opposite, the difference in periods cannot be explained by such  selection biases, but it really shows the power of space observations in the long period range.

We tested whether the K2 long cadence (30-min) sampling has any effect on the period (and amplitude) determination. Since we considered rotational periods that are longer the 2 hours (i.e. below the Nyquist frequency-limit), we expect that despite the long integration, we still can retrieve the periods. This is indeed the case, as is demonstrated by a series of Monte-Carlo simulation. In the course of these simulations we computed that in what fraction of the cases our algorithm can find the right (injected) period, which was simulated by a pure sinus wave. In these tests the photometric noise (depending on the brightness), the observation length, the duty cycle, and the rotational period were varied. We found that our sensitivity does not decrease in the relevant period range (>2h) due to the long integration times. The most important factors are instead the length of observations and the noise level. With the long cadence sampling the measured amplitude decreases with respect to the true variation for short period signals, but our simulations show that we do not loose asteroids due to this effect. The worst case scenario is a decrease of the amplitude by 37\% at a rotation period of 2 hours, which is still detectable in our K2 data in the case of an assumed 0.1 magnitude true variation down to the 20th magnitude.

\section{Conclusions}

\begin{figure}
\centering
\includegraphics[width=6.1cm, angle=270]{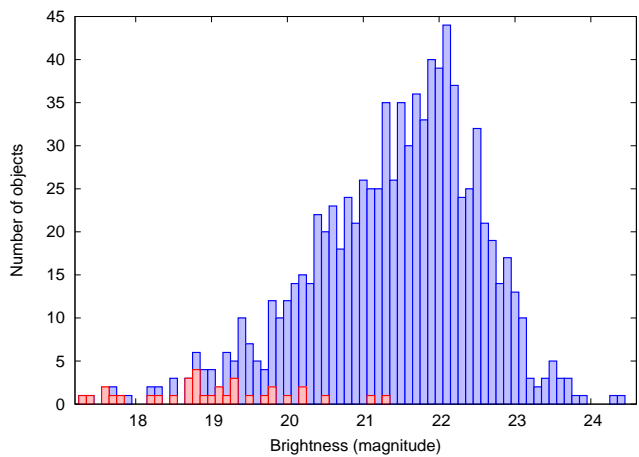}
\caption{Magnitude distribution of the   asteroids in the M35 field seen by \textit{Kepler} (924, blue columns), and the selected sample where significant rotational signal (period or trend) could be derived (35, red columns). Two brighter objects were omitted from the figure for the sake of clarity.}
\label{magdist}
\end{figure}

We utilized the \textit{Kepler} space telescope for the first time to derive quasi-continuous light curves of a large number of main-belt asteroids in long cadence mode (29.4~min sampling). The main conclusions of this work are the following:

\begin{itemize}

\item{Out of 924 (96) asteroids in the M35 (Nereid) field in Campaign~0 (3), 867 (88) had twelve or more useful photometric data points and only 23 (14) exhibited clear periodicities which is attributed to rotation. In addition, 12 (0) objects showed a slow trend or were observed through an incomplete rotational cycle implying a long rotational period.}

\item{By comparing the M35 and Nereid samples we found a remarkable difference regarding the number of main-belt asteroids with detected rotational periods in the two fields.
While in the dense M35 field only 4.0\% of the asteroids showed clear periodicities or trend, in the Nereid field we recovered periodicities in 15.9\% of the observed asteroids. The difference is significant given the large number of observed asteroids in both fields. To explain this difference we propose two arguments: }

\item{First, we conclude that the dense stellar field precluded the derivation of meaningful photometry in the case of many asteroids, because too many points had to be discarded, due to the disturbing effects of stellar residuals along the paths of the asteroids (see  Fig.~\ref{duty}). This is partly explained by the undersampled PSFs delivered by the {\it Kepler} spacecraft.}

\item{Second, in Fig.~\ref{magdist} we plot the magnitude distribution of our full M35 asteroid sample (in blue) as seen from Kepler, and also those that showed periodicities or long-term trends in their light variations (red). The plot convincingly shows that our sample is heavily dominated by faint targets (>20~mag). Together with Fig.~\ref{magerr} this clearly demonstrates that there is a rather low chance to pull out rotational signal of asteroids below the ${\rm 20^{th}}$ magnitude brightness limit. This  reasoning helps to explain the relatively low rate of recovered rotational periods in our fields, especially in the M35 superstamp.}

\item{More sophisticated photometric methods
may improve our results and provide more reliable and robust space photometric data of moving objects. Testing of such methods is currently is under way.}

\end{itemize}

\begin{table}
\centering
\caption{Rotational signal detected in   asteroids observed in the K2 M35 superstamp.}
\begin{tabular}{ccccc}
\hline\hline
ID & period & ampl. &  ref. \\
& [h] &  [mag] &  \\
\hline
228       &   6.437 $\pm$ 0.047   &   0.158  &    this paper \\
          &   6.47     &   0.27   &    \citet{ivarsen2004} \\
          &   6.484    &   0.27   &   \citet{cooney2005} \\
767       &   >60.     &   > 0.1  &   this paper\\
2785      &   5.479 $\pm$ 0.031   &   0.310  &    this paper \\
          &   5.49     &   0.45   &    \citet{polishook2009} \\
          &   5.478    &          &    \citet{hanus2016} \\
3345      & >34.       &   >0.7   &    this paper \\
          & 187.       &   0.59   &    \citet{benishek2014} \\
3512      &   6.67  $\pm$ 0.25  &   0.212  &    this paper \\
          &   6.782    &   0.35   &    \citet{ditteon2004} \\
          &   6.784    &   0.25   &    \citet{skiff2011} \\
3903      &  28.09 $\pm$ 0.97   &   0.190  &    this paper \\
4282      & >34.       &   >0.6   &    this paper \\
7122      & >96.       &  >0.6    &    this paper \\
7241      &  59.6 $\pm$ 5.7   &   0.606  &    this paper \\
10269     &  10.81 $\pm$ 0.33   &   0.090  &    this paper \\
10683     &  75.3 $\pm$ 3.1  &   1.0    &    this paper \\
12379     &  50.6 $\pm$ 3.1   &   0.378  &    this paper \\
13243     &  88.4 $\pm$ 9.8   &   0.784  &    this paper \\
14439     &   9.94 $\pm$ 0.38   &   0.322  &    this paper \\
17771     &  15.97  $\pm$ 0.26  &   0.456  &    this paper \\
21207     & >120.      &  >0.5    &    this paper \\
24192     &  19.9 $\pm$ 1.0   &   0.464  &    this paper \\
25468     &   5.573 $\pm$ 0.078   &   0.482  &    this paper \\
29296     & >48.       &   >0.9   &    this paper \\
30329     &  28.4 $\pm$ 2.1   &   0.708  &    this paper \\
37750     &   5.15 $\pm$ 0.21   &   0.204  &    this paper \\
42573     &   3.888 $\pm$ 0.083   &   0.164  &    this paper \\
          &   3.887    &   0.55   &    \citet{waszczak2015} \\
49039     & >72.       &  >0.8    &    this paper \\
49193     &   5.66  $\pm$ 0.55  &   0.400  &    this paper \\
52786     & >144.      &  >0.5    &    this paper \\
55949     &  11.16 $\pm$ 0.15   &  0.324   &   this paper \\
          &  11.135    &   0.43  & \citet{waszczak2015} \\
57648     &   8.359 $\pm$ 0.079   &   0.266  &    this paper \\
66340     & >62.       &  >1.0    &    this paper \\
67087     &  17.7 $\pm$ 1.5   &   0.316  &    this paper \\
69759     & >41.       &  >0.6    &    this paper \\
69853     &   8.17 $\pm$ 0.83   &   0.454  &    this paper \\
109978    & >144.      &  >0.3    &    this paper \\
115554    &   8.11 $\pm$ 0.91   &   0.402  &    this paper \\
149686    & >72.       &  >1.0    &    this paper \\
218609    &   9.72 $\pm$ 0.22   &   0.690  &    this paper \\
\hline
\end{tabular}
\label{tab5}
\end{table}

\begin{table}
\centering
\caption{Observed periods and amplitudes of the asteroids in the K2 Nereid field.}
\begin{tabular}{cccc}
\hline\hline
ID & period & ampl. & ref. \\
& [h] & [mag] & \\
\hline
2954      &   4.691 $\pm$ 0.010  &  0.174  &  this paper \\
          &   4.690     &  0.21   &  \citet{wisniewski1997} \\
          &   4.68      &  0.25   &  \citet{ferrero2010} \\
3785      &   3.782 $\pm$ 0.007    &  0.256  &  this paper \\
          &   3.7992    &  0.30   &  \citet{behrend2009} \\
9105      &   4.769 $\pm$ 0.032    &  0.770  &  this paper \\
31907     &   3.786  $\pm$ 0.081   &  0.076  &  this paper \\
57575     &   4.641 $\pm$ 0.077    &  0.436  &  this paper \\
87632     &  14.6 $\pm$ 2.7    &  0.128  &  this paper \\
211339    &   5.14 $\pm$ 0.20    &  0.734  &  this paper \\
242602    &   8.600 $\pm$ 0.094    &  0.706 &  this paper \\
249866    &   3.71 $\pm$ 0.20    &  0.172   &  this paper \\
          &   3.7982    &  0.18   &  \citet{waszczak2015} \\
250648    &   5.95 $\pm$ 0.17    &  0.564  &  this paper \\
311247    &  11.94 $\pm$ 0.48    &  0.584  &  this paper \\
314616    &   7.81 $\pm$ 0.24    &  0.270  &  this paper \\
\hline
\end{tabular}
\label{tab6}
\end{table}

\begin{acknowledgements}
This project has been supported by the 
Lend\"ulet LP2012-31 Young Researchers Program,
the Hungarian OTKA grants K-109276 and K-104607,
the Hungarian National Research, Development and Innovation Office (NKFIH) grants K-115709 and PD-116175, the GINOP-2.3.2-15-2016-00003 grant, 
and by City of Szombathely under agreement no. S-11-1027. 
The research leading to these results has received funding from the 
European Community's Seventh Framework Programme (FP7/2007-2013) under 
grant agreement no. 312844 (SPACEINN), the European Union’s Horizon 2020 Research and Innovation Programme, under Grant Agreement no 687378 (SBNAF),  and the ESA 
PECS Contract Nos. 4000110889/14/NL/NDe and 4000109997/13/NL/KML, and the European Union’s Horizon 2020 Research and Innovation Programme, Grant Agreement no 687378. 
Gy.~M.~Sz., Cs.~K. and L.~M. were supported by the J\'anos Bolyai Research 
Scholarship. Funding for the K2 spacecraft is provided by the NASA Science 
Mission directorate. The authors thank the  hospitality of the Veszprém Regional Centre of the Hungarian Academy of Sciences (MTA VEAB), where part of this project was carried out.
We acknowledge the \textit{Kepler} team and engineers for their efforts to keep this fantastic instrument alive and for allocating the large pixel mosaics. 
All of the data presented in this paper were obtained from the 
Mikulski Archive for Space Telescopes (MAST). 
STScI is operated by the Association of Universities for Research in 
Astronomy, Inc., under NASA contract NAS5-26555. Support for MAST for 
non-HST data is provided by the NASA Office of Space Science via 
grant NNX13AC07G and by other grants and contracts. We thank the referee for his/her useful comments which helped to improve the paper significantly.
\end{acknowledgements}

\end{document}